\documentclass[%
 reprint,
nofootinbib,
 amsmath,amssymb,
 aps,onecolumn
]{revtex4}

\usepackage{amssymb,amsmath,amsfonts,amsbsy,graphicx}
\usepackage{color}
\usepackage{psfrag}
\usepackage{stackrel}

\newcommand\be{\begin{equation}}
\newcommand\ee{\end{equation}}
\newcommand\ba{\begin{eqnarray}}
\newcommand\ea{\end{eqnarray}}\newcommand\eq{\begin{equation}}           
\newcommand\en{\end{equation}}

 \newcount\colveccount
\newcommand*\colvec[1]{
        \global\colveccount#1
        \begin{pmatrix}
        \colvecnext
}
\def\colvecnext#1{
  #1
  \global\advance\colveccount-1
        \ifnum\colveccount>0
                \\
                \expandafter\colvecnext
        \else
                \end{pmatrix}
        \fi
}
\input epsf

\usepackage{ulem}
\usepackage{amssymb}
\usepackage{amsmath}
\usepackage{bm}
\usepackage{graphicx}
\usepackage{color}
\usepackage{subfig}
\usepackage{caption}

\def\gsim{\;\rlap{\lower 2.5pt
 \hbox{$\sim$}}\raise 1.5pt\hbox{$>$}\;}
\def\lsim{\;\rlap{\lower 2.5pt
 \hbox{$\sim$}}\raise 1.5pt\hbox{$<$}\;}

\usepackage{graphicx}
\usepackage{url}
\usepackage{amssymb,amsmath}
\usepackage{amsbsy}

\usepackage{graphicx}
\usepackage{url}
\usepackage{amssymb,amsmath}
\usepackage{amsbsy}

\begin{document}
\title{
  Probing axion dark matter with 21cm fluctuations from minihalos
}
\author{Kenji Kadota$^1$, Toyokazu Sekiguchi$^2$, Hiroyuki Tashiro$^3$ \\
  {\small $^1$ Center for Theoretical Physics of the Universe, Institute for Basic Science (IBS), Daejeon, 34051, Korea} \\
  {\small $^2$ Institute of Particle and Nuclear Studies, KEK, 1-1 Oho, Tsukuba 305-0801, Japan} \\
  {\small $^3$ Department of physics and astrophysics, Nagoya University, Nagoya 464-8602, Japan}
  }

\begin{abstract}
  If the symmetry breaking inducing the axion occurs after the inflation, the large axion isocurvature perturbations can arise due to a different axion amplitude in each causally disconnected patch. This causes the enhancement of the small-scale density fluctuations which can significantly affect the evolution of structure formation. The epoch of the small halo formation becomes earlier and we estimate the abundance of those minihalos which can host the neutral hydrogen atoms to result in the 21cm fluctuation signals. We find that the future radio telescopes, such as the SKA, can put the axion mass bound of order $m_a \gtrsim 10^{-13}$ eV for the simple temperature-independent axion mass model, and the bound can be extended to of order $m_a \gtrsim 10^{-8}$eV for a temperature-dependent axion mass.
\end{abstract}

\maketitle
\setcounter{footnote}{0} 
\setcounter{page}{1}\setcounter{section}{0} \setcounter{subsection}{0}
\setcounter{subsubsection}{0}

\section{Introduction}
There can exist many symmetries in the early Universe even though they are broken in the current Universe, and it would be of great interest to seek the signals associated with those symmetries to unveil the nature of the Universe's evolution. An intriguing possibility is the existence of pseudo-Nambu-Goldstone bosons (pNGB) which can arise in the early Universe when a continuous global symmetry is spontaneously broken.
An example includes the Peccei-Quinn (PQ) symmetry which was introduced to solve the strong CP problem in the QCD, and the axion arises as a pNGB which can also be a potential dark matter candidate. More generally, the axion-like particles where the mass and coupling are treated independently have gained the growing interests for their cosmological implications such as the small scale structure formation by the fuzzy dark matter \cite{Peccei:1977hh,Weinberg:1977ma,Wilczek:1977pj,Preskill:1982cy,Abbott:1982af,Dine:1982ah,sik1983,raf1987,hu2000,hw2009,arv2009,as2009,kad2015,2017PhRvL.119c1302I,ana2017,Kadota:2013iya,kel2017,hua2018,hook2018,kad2015b,har1992,fed2019,kad2019,shi2019,Bauer:2020zsj,fai2017}. We call such a U(1) global symmetry the PQ symmetry and the associated pNGB the axion in this paper, and we study the effects of the axion dark matter on the 21cm fluctuations when the PQ symmetry is broken after the inflation.

A characteristic feature of the post-inflationary PQ symmetry breaking scenarios is the existence of the large isocurvature perturbations. Even though there are tight bounds on the isocurvature perturbations from the current cosmological observables such as the CMB and Lyman-$\alpha$ forest, there still remains a room for the large isocurvature component at small scales which can affect the formation of small halos \cite{hog1988,kolb1994,zu2006,Ringwald:2015dsf,marsh15,sh1994,tk1998,Feix:2019lpo,Feix:2020txt,Irsic:2019iff,Shimabukuro:2020tbs,Afshordi:2003zb,Murgia:2019duy,Kadota:2020ahr}.
We discuss the 21cm probes on such isocurvature perturbations by the forthcoming radio surveys such as the SKA (Square Kilometer Array) \cite{skawebpage}. The hydrogen is the most abundant element in the Universe and the 21cm emission/absorption signals due to the neutral hydrogen can offer us a promising probe on the small scales far beyond those currently accessible by the other means. In particular, the minihalos are the biased tracers of the underlying matter density fluctuations and they can host the dense neutral hydrogen atoms to enhance the 21cm signals \cite{Shapiro:1998zp,il2002,Iliev:2002ms}. An advantage of the use of minihalos is their small sizes which enables us to probe the scales down to of order $k\sim 10^3/$Mpc, and the applications to probe such small scales include the studies of the warm dark matter and the primordial non-Gaussianity besides the axions in the post-inflationary PQ breaking scenarios to be discussed in this paper \cite{Sekiguchi:2013lma,Sekiguchi:2014wfa,Takeuchi:2013hza,Sekiguchi:2017cdy,Chongchitnan:2012we,Sekiguchi:2018kqe}. While the current bounds on the axion mass are of order $m_a\gtrsim 10^{-20}$ eV using the CMB and $m_a\gtrsim 10^{-17}$ eV from the Lyman-$\alpha$ forest for the temperature independent axion mass scenarios \cite{Feix:2019lpo,Feix:2020txt,Irsic:2019iff}, our study on the 21cm fluctuations from the small halos predicts the extension of those sensitivities to $m_a\gtrsim 10^{-13}$eV.

We first discuss the nature of isocurvature perturbations in the post-inflationary PQ symmetry breaking scenarios in \S\ref{pqmodel}. \S\ref{21cmmodel} outlines the formalism to estimate the 21cm angular power spectrum from the minihalos. \S\ref{fisher} presents the likelihood analysis by the Fisher matrix calculations and \S\ref{axionmass} discusses the corresponding axion mass parameter ranges the future 21cm fluctuation signals can be sensitive to, followed by the discussion/conclusion in \S\ref{disconc}.

\section{Isocurvature density fluctuations in post-inflationary PQ symmetry breaking scenarios}
\label{pqmodel}
We study the scenarios where the PQ symmetry breaking occurs after the inflation, for which the characteristic feature is the existence of large isocurvature perturbations dominating the conventional adiabatic perturbations at small scales. 
We consider the potential for a PQ scalar field $\phi$ given by $V(\phi)=\lambda (|\phi|^2 -f_a^2/2)^2$ for which $\phi$ settles down at a minimum $|\phi|=f_a/\sqrt{2}$ when the global U(1) symmetry is spontaneously broken. We assume each causally disconnected horizon patch possesses a randomly distributed phase $\theta \ni [-\pi,\pi]$ for $\phi=|\phi|e^{i\theta}$ and identify the angular component $a\equiv f_a \theta$ as an axion with an axion decay constant $f_a$. The axion later can obtain the mass $m_a$ non-perturbatively and this leads to the axion potential $m_a^2 a^2/2$. Because of the randomly distributed $\theta$, the energy density ($\propto \theta^2$) in each horizon patch is different and can lead to the large density perturbations over different horizons. The perturbations are smoothed out inside the horizon scale due to the gradient term in the Lagrangian (Kibble mechanism \cite{kib1976}), so that we model the initial axion isocurvature perturbations when the axion starts oscillation as the white noise power spectrum with the comoving horizon scale $k_{osc}=a_{osc}H_{osc}$ as a cutoff scale 
     \ba
     P_{iso}(k,t_{osc})= P_0\Theta(k_{osc}-k), P_0=\frac{24}{5} \frac{\pi^2}{k_{osc}^3}
     \ea
     where $\Theta$ is the Heaviside function. The normalization factor $P_0$ is from the relation for the variance $\sigma^2 \equiv \langle \delta_a^2 \rangle = (2\pi)^{-3} \int P(k) d^3 k$ where the initial axion density is of order unity $ \langle \delta_a^2 \rangle=4/5 $ ( $\delta_a \equiv (\rho_a-\bar{\rho}_a)/ \bar{\rho}_a$) for a randomly distributed $\theta$ (we used $\rho_a \propto \theta^2, \langle \theta^2 \rangle=\pi^2/3 $).

The total matter power spectrum we consider in our study is hence the sum of the conventional adiabatic perturbations (presumably originated from the inflation) and the axion isocurvature perturbations \cite{2020AJ....159...49D,fai2017}
\ba
P(k,z)=P_{ad}(k) D^2(z)+
\Theta(k_{osc}-k)
\frac{24\pi^2}{5k_{osc}^3}
\left(
\frac{D(z)}{D(z_{*})}
\right)^2
\left(
\frac{1+z_{eq}}{1+z_{*}}
\right)^2
\label{matterp}
\ea
where $D$ is the growth factor, $z_*$ is an arbitrary redshift deep in the matter domination epoch and $z_{eq}$ is the redshift at the matter-radiation equality. The axion starts oscillation (when the axion mass becomes comparable to the Hubble scale) during the radiation dominated epoch in our scenarios. The small logarithmic growth of fluctuations during the radiation epoch does not affect our discussions, and we approximate  the axion density fluctuation amplitude at the matter-radiation equality as that when the axion starts oscillation  \cite{hog1988,zu2006,har2016,fai2017} $^1$.\footnotetext[1]{
  Even though the initial isocurvature fluctuations are of order unity in our scenarios, it would be also worth exploding whether or not the initial fluctuations can exceed far beyond order unity in a realistic axion model (see for instance Ref. \cite{kolb1994,ena2017}).}
We assume the axion constitutes the whole cold dark matter in the Universe unless stated otherwise (we briefly discuss the partial axion dark matter scenarios at the end of the paper).

\begin{figure}[htbp]
  
   \begin{tabular}{c}
                 \includegraphics[width=0.8\textwidth]{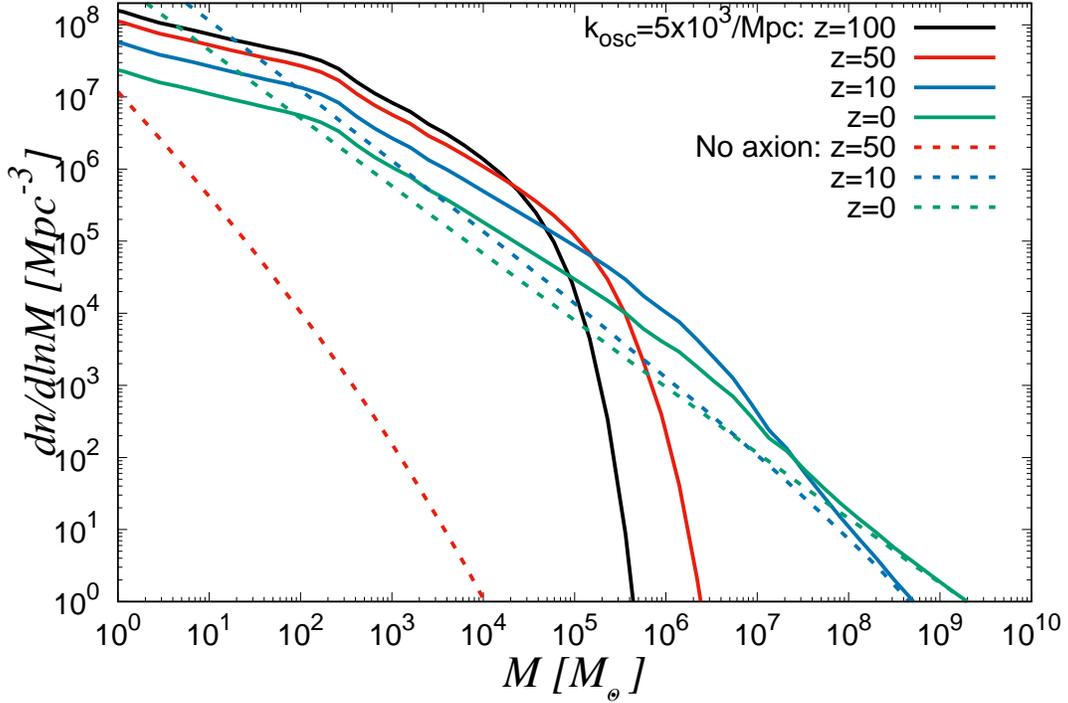}

    \end{tabular}

    
   \caption{The halo mass functions $dn/dlnM$ [Mpc$^{-3}]$ as a function of the halo mass $M[M_{\odot}]$ for the adiabatic plus axon isocurvature perturbations ($k_{osc}=5000/$Mpc) and those for the adiabatic perturbations alone (for which $z=100$ case is too small to be shown in this figure). 
   }

 %
   \label{propermasscutk5000}
\end{figure}

In the presence of such large isocurvature perturbations at small scales, the formation of small structures can occur earlier as illustrated in Fig. \ref{propermasscutk5000} which shows the mass function (the abundance of halos as a function of the halo mass) \cite{1999MNRAS.308..119S}.
The minihalos of our interest are the virialized halos with the mass ${\cal O}(10^4 \sim 10^7) M_{\odot}$ (the more precise redshift dependent mass range to be given in the next section)
which are filled with the neutral hydrogen atoms. The halo abundance for our axion scenarios is much bigger than that for the scenarios without axions at $z=100$ (the abundance for no axion case at $z=100$ is too small to be shown in this figure) because the isocurvature perturbations can dominate the adiabatic perturbations at a small scale and those small halos are produced much earlier than those for the conventional adiabatic scenarios. At $z=50$, for the axion scenarios, the small mass halo abundance is already saturated in the hierarchical structure formation process while the merging of those small halos into the larger ones can increase the larger mass halo abundance (and hence decrease the small halo abundance compared with that at $z=100$). 
At $z=10$, the minihalo abundance for the axion scenarios still exceeds that for the adiabatic scenarios for the minihalo mass range of our interest, while the abundance of larger halos for adiabatic perturbation scenarios match with that for the axion scenarios because the adiabatic perturbations dominate the isocurvature perturbations at large scales.

\begin{figure}[htbp]

   \begin{tabular}{c}
                   \includegraphics[width=0.8\textwidth]{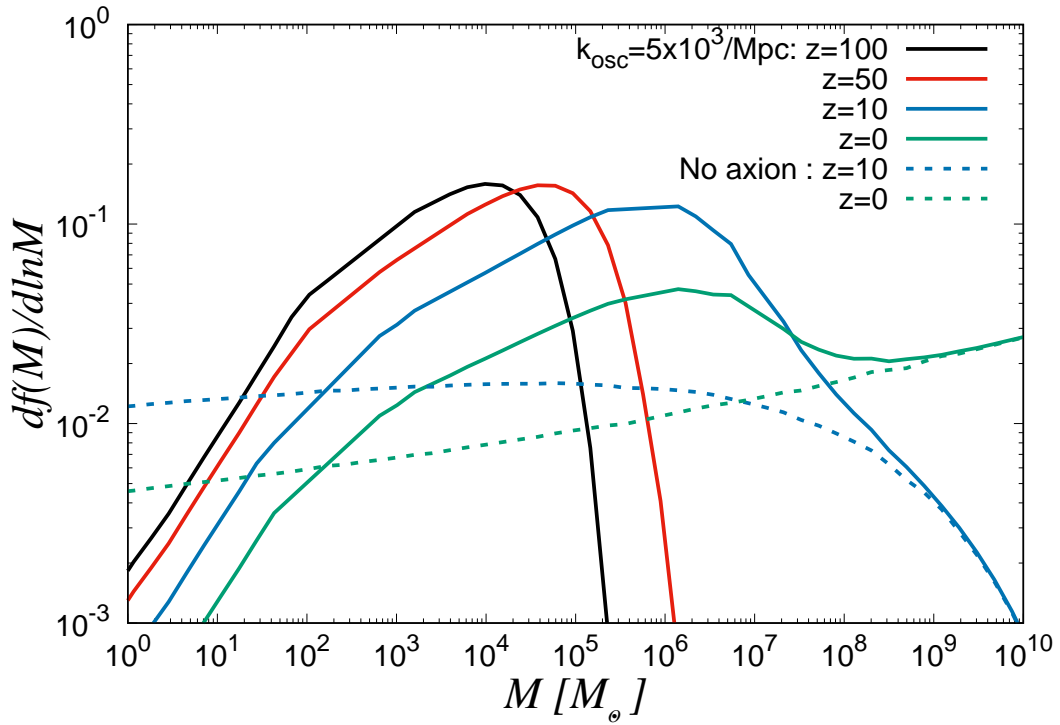}
    \end{tabular}

    
   \caption{
The mass fraction of isolated collapsed halos $df(M)/d\log M$ using the Press-Schechter formalism. The mass fractions for the adiabatic perturbations alone at $z=100, 50$ are too small to be shown in this figure. 
   }

   \label{massfractionsmoothNov30}
\end{figure}

It would be also illustrative to see the mass fraction which reside in the collapsed halos. According to the Press-Schechter formalism, the mass fraction and mass function are related by $\frac{df}{d\ln M }=\frac{dn}{d\ln M} \frac{M}{\bar{\rho}}$ ($\bar{\rho}$ is the background density). We can see from Fig. \ref{massfractionsmoothNov30} that the large fraction of the mass is already locked in the collapsed halos even at $z=100$ which is in stark contrast to the no-axion scenarios. For the minihalo mass range of our interest, the mass fraction in the collapsed objects is about an order of magnitude bigger than that in the no-axion scenarios. The actual difference can be even bigger because the Press-Schechter formalism cannot take account of subhalo abundance. Many minihalos were produced at the high redshifts and consequently possess the much higher densities than the no-axion scenarios because the densities of halos are proportional to $(1+z_f)^3$ depending on the formation redshifts $z_f$. Hence those denser minihalos are more resilient to the tidal disruptions and the actual minihalo abundance could be bigger than the abundance of the isolated minihalos estimated by the Press-Schechter formalism. The detailed numerical simulations to study the survival probabilities of those dense minihalos in the axion cosmology are left for the future work.

As a promising probe on the small scale structures at a high redshift, we discuss the 21cm fluctuation signals form the minihalos at a high redshift (before the completion of reionization $z\gtrsim 6$) in the following sections.

\section{21cm angular power spectrum from minihalos}
\label{21cmmodel}
We first briefly review the estimation of 21cm fluctuations from minihalos \cite{Shapiro:1998zp,il2002,Sekiguchi:2017cdy}. The minihalos of our interest are the virialized halos of dark and baryonic matter with the mass range ${\cal O}(10^4)M_{\odot} \lesssim M_{halo} \lesssim {\cal O}(10^7) M_{\odot}$ which are filled with the neutral hydrogen atoms. The additional power at small scales leads to the earlier minihalo formation as discussed in the last section. The neutral hydrogen atoms in the minihalos can be hot and dense enough to emit the observable 21cm line spectra. We study the angular fluctuations in this 21cm background to probe the small-scale structures of the Universe.
The minimum and maximum minihalo masses of our interest are redshift dependent, and we use, for the minimum mass, the baryon Jeans mass \cite{2001PhR...349..125B}
\ba
M_{min}(z)=
5.7 \times 10^3
\left(
\frac{\Omega_m h^2}{0.15}
\right)^{-1}
\left(
\frac{\Omega_b h^2}{0.02}
\right)^{-3/5}
\left(
\frac{1+z}{10}
\right)^{3/2}
M_{\odot}
\label{mminhalo}
\ea
and the maximum mass is
\ba
M_{max}(z)
=
3.95 \times 10^7
\left(
\frac{\Omega_m h^2 }{ 0.15}
\right)^{-1/2}
\left(
\frac{1+z}{10}
\right)^{-3/2}
M_{\odot}
\label{mmaxhalo}
\ea
which corresponds to the virial temperature $T_{vir}=10^4$ K below which the atomic cooling is inefficient for the star formation \cite{2001PhR...349..125B,Iliev:2001he,il2002} $^2$. \footnotetext[2]{We leave a more detailed estimation for the relevant halo mass range for the future work. It would require the numerical simulations taking account of the reionization and the radiative feedback, which for instance could enhance the baryon Jeans mass. We also point out that the axion dark matter Jeans scale inside which the quantum pressure prevents the fluctuation growth (corresponding to the de Broglie length scale of the axion dark matter) is smaller than the cutoff scale $k_{osc}$ of our interest and hence can be safely neglected in our analysis \cite{hu2000,hw2009,arv2009,shi2019,Kadota:2013iya,2017PhRvL.119c1302I,fai2017}.} 
The minihalos of our interest are hence not large enough to host a galaxy or even stars which can source the ionization because we are interested in the neutral hydrogen. 

The brightness temperature along the line of sight with a distance $r$ from the halo center reads
\ba
T_b(r)=T_{CMB} e^{-\tau(r)}+\int^{\tau(r)}_{0} T_S e^{-\tau'} d\tau'
\ea
where $\tau(r)$ is the optical depth of neutral hydrogen through the halo, and $T_{CMB},T_S$ are the CMB temperature and the spin temperature inside the halo \cite{1959ApJ...129..525F,1997ApJ...475..429M,2002ApJ...579....1F}.
The collisional excitation of the neutral hydrogen can be efficient inside the virialized halo where the gas can be nonlinear and hot enough for the gas collisions to dominate the excitation by the CMB photons. In our scenarios under consideration, therefore, the spin temperature is strongly coupled with the gas kinetic temperature exceeding the CMB temperature, $T_S\approx T_{k} > T_{CMB}$ \cite{Shapiro:1998zp,Iliev:2001he}. 
\\
We here derive the estimation for $\overline{T}_b$, the averaged brightness temperature from all the minihalos of our interest.
$T_b$ is related to the specific intensity of a blackbody radiation in the Rayleigh-Jeans limit (the relevant frequencies are much smaller than the peak frequency of the CMB blackbody) by the relation $^3$\footnotetext[3]{In the radio astronomy, we conventionally use the brightness temperature $T_b$ instead of the observed intensity $I_{\nu}=dF/d\nu d\Omega$ (the flux per unit frequency per solid angle).}
\ba
\frac{dF}{d\nu d\Omega}=\frac{2\nu^2 }{c^2} k_B T_b(\nu)
\ea
The line-integrated flux $F$ from a halo is obtained by the flux calculated for $\nu=\nu_0$ ($\nu_0=1.4$ GHz is the 21cm line in a rest frame) multiplied by the effective line-width and the solid angle
\ba
F=\int d\nu d\Omega \frac{dF}{d\nu d\Omega}
=
\Delta \Omega  \Delta \nu_{eff} \left. \frac{dF}{d\nu d\Omega}\right\vert_{\nu_0}=\Delta \Omega \Delta \nu_{eff}  \frac{2 \nu_0^2 }{c^2}k_B T_{b,\nu_0}
\ea
where $\Delta \Omega=A/D_A^2$ is the solid angle subtended by a halo (the geometric cross section $A=\pi r_{halo}^2$ for a halo of mass $M$ and radius $r_{halo}$), the angular diameter distance $D_A=D_c/(1+z)$ ($D_c$ is the comoving distance), $\Delta \nu_{eff}=[\phi(\nu_0) (1+z)]^{-1}$ is the effective redshifted line width with the thermal Doppler broadened line profile $\phi(\nu)=(\Delta \nu \sqrt{\pi})^{-1} \exp\left(  -[ (\nu-\nu_0)/\Delta \nu]^2 \right) $ and $\Delta \nu =(\nu_0/c) \sqrt{2k_B T_{gas}(z)/m_H}$ \cite{Chongchitnan:2012we}.


The averaged intensity from all the minihalos within a given solid angle and a given frequency interval is
\ba
\overline{
  \frac{dF}{d\nu d\Omega}} d\nu d\Omega
  =
  \int
  dM
  \frac{dn}{dM}
  \langle
  F
  \rangle
  dV
  \ea
  where $dV=D_c^2 dD d\Omega$ is the differential comoving volume and $\langle F \rangle$ represents the flux averaged over the geometric cross section of a halo $A$
  \ba
  \langle
  F
  \rangle=
  \frac{1}{A}
  \int dA F
  \ea
  We hence obtain the average differential flux per frequency per solid angle
  \ba
  \overline{
  \frac{dF}{d\nu d\Omega} } 
  =
  \frac{c(1+z)^2}{\nu_0 H(z)} D_c^2 \int dM \langle F \rangle \frac{dn}{dM}
  \ea
  where we used $dD_c/dz=c/H(z), |d\nu/dz|=\nu_0/(1+z)^2$.
  We then finally obtain the averaged brightness temperature \cite{il2002}
  \ba
  \overline{T}_b=
  \frac{c^2}{2\nu_0^2 k_B}
    \overline{
      \frac{dF}{d\nu d\Omega} }
    =
    \frac{c(1+z)^4}{\nu_0 H(z)} \int^{M_{max}}_{M_{min}}   \Delta \nu_{eff} \langle T_{b,\nu_0}\rangle A \frac{dn}{dM}dM
      \ea
      The properties of the minihalo such as the gas density profile determine $A, \phi(\nu_0), \langle  T_b \rangle$ to give us the estimation of 21cm emission flux. We follow the calculations in the truncated isothermal sphere given in Refs \cite{Shapiro:1998zp,Iliev:2001he} where the minihalo profile is modeled by a non-singular truncated isothermal sphere in virial and hydrostatic equilibrium and the internal structure of a halo is characterized by the total mass $M$ and the collapse redshift. 
The observed differential brightness temperature from a halo with respect to the CMB is
\ba
\Delta T_b=\frac{\langle T_b \rangle -T_{CMB}(z)}{1+z}
\ea
where $z$ is the redshift at the emission from a minihalo, and the corresponding averaged brightness temperature follows from the above derivation as
  \ba
\overline{\Delta T}_b
    =
      \frac{c(1+z)^4}{\nu_0 H(z)} \int^{M_{max}}_{M_{min}}   \Delta \nu_{eff} \langle  \Delta T_{b,\nu_0}\rangle A \frac{dn}{dM}dM
  \ea

  

      
The minihalos are the tracers of the underlying matter distribution, and the minihalo clustering can be related to the underlying density fluctuations through the bias factor. We hence apply the conventional bias formalism based on the halo models \cite{Cooray:2002dia} to express the 21cm line fluctuations (fluctuations in the differential brightness temperature) from the minihalos in the direction $\hat{n}$ as
 \ba
 \delta T_{21cm}
 =   \overline{\Delta T_b}(z)
 \bar{b}(z)
     \delta(\vec{x}=r(z)\hat{n},z)
\ea
where $\delta(\vec{x},z)$ is the matter density fluctuations 
 at the comoving coordinate $\vec{x}$ with the comoving distance $r(z)$ from us to the redshift $z$. $\bar{b}(z)$ is the flux-weighted effective bias averaged over the mass function 
      \ba
      \bar{b}(z)
      =
      \frac{\int ^{M_{max}}_{M_{min}} dM \frac{dn}{dM} {\cal F} (z,M) b(M,z)}
{\int^{M_{max}}_{M_{min}} dM \frac{dn}{dM} {\cal F} (z,M)}
\ea
where ${\cal F}(= \langle \delta T_b \rangle A \sigma_V$ with the velocity dispersion of a minihalo $\sigma_V$) is the flux from a minihalo \cite{Shapiro:1998zp,Shapiro:2005cx,Chongchitnan:2012we} and $b(M,z)$ is the halo bias of Ref. \cite{Mo:1995cs}.

\begin{figure}
  
   \begin{tabular}{c}

                    \includegraphics[width=0.8\textwidth]{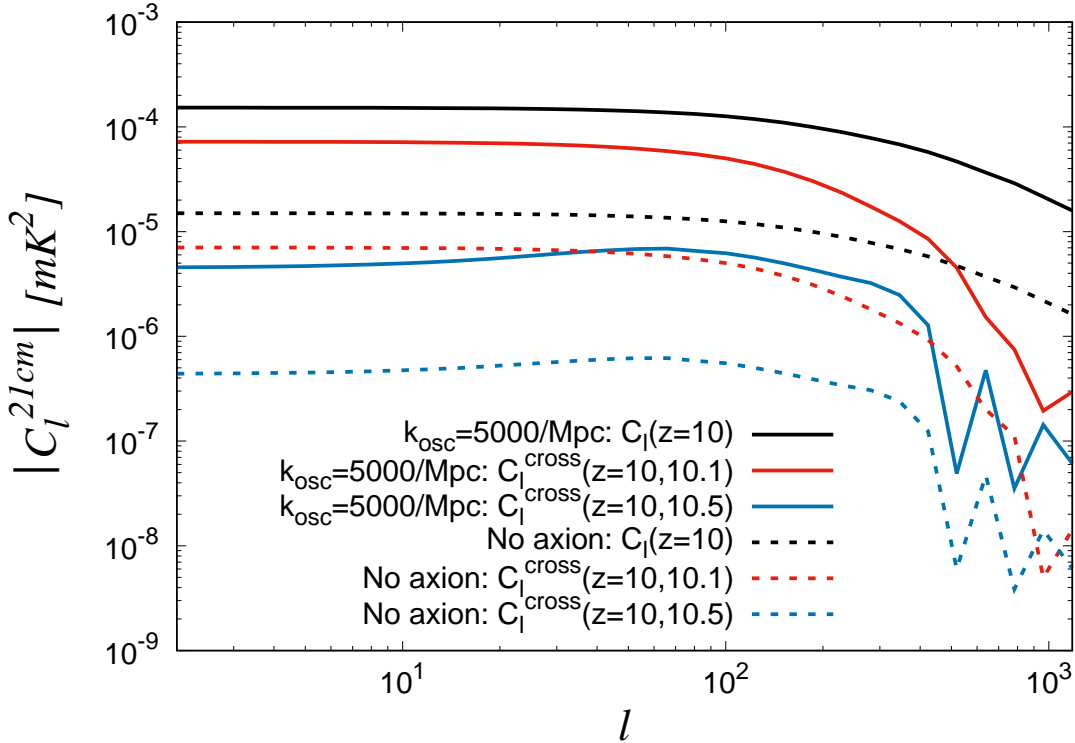}
    \end{tabular}

    
   \caption{The 21cm angular power spectra at $z=10, C_l(z=10)$, and the cross power spectra $C_l^{cross}(z_1,z_2)$ between the redshift bins $z=z_1$ and $z=z_2$ (two examples of $(z_1,z_2)=(10, 10.1)$ and $(10,10.5)$ are shown). The absolute values $|C_l|$ are shown in this log plot. The cross power spectrum amplitudes decrease for the redshift binds with a larger redshift difference.
     }

   \label{cell21z10}
\end{figure}

\begin{figure}[htbp]
  
   \begin{tabular}{c}

               \includegraphics[width=0.8\textwidth]{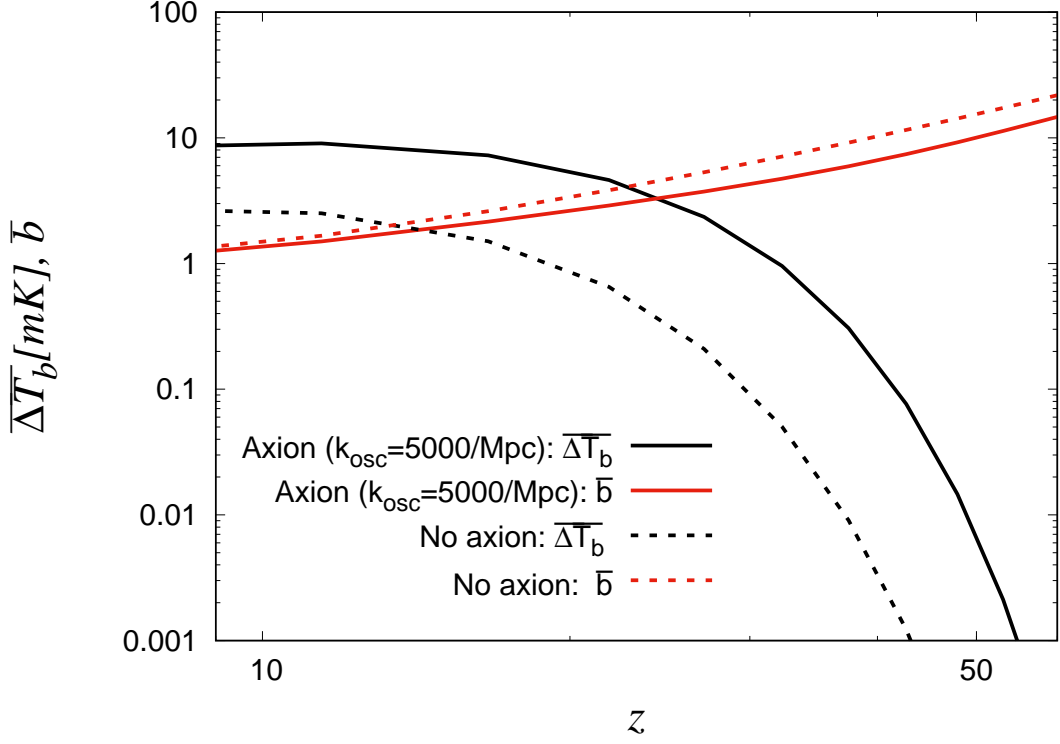}
    \end{tabular}

    
   \caption{
 $\overline{\Delta T_b}(z)$[mK] and $\bar{b}(z)$ for the adiabatic plus axion isocurvature perturbations $(k_{osc}=5\times 10^3/$Mpc) and for no axion (only adiabatic perturbations) scenarios. 
   }

   \label{figtbbias}
\end{figure}

      We expand the brightness temperature fluctuations in terms of the spherical harmonics with the multipole moments
      \ba
      a_{lm}^{21cm}(z)=
      \int d\hat{\bf{n}} \delta  T_{21cm}(\hat{\bf{n}},z)  Y^{m*}_l(\hat{\bf{n}})
      \ea
and the angular power spectrum is
\ba
C_l^{21} (z,z')= \langle
a_{lm}^{21}(z) a_{lm}^{21*} (z')
\rangle
=
\overline{\Delta T_b}(z)
\overline{\Delta T_b}(z')
\bar{b}(z)
\bar{b}(z')
C_l^{matter}
\ea
\ba
C_l^{matter}
(z,z')
=D(z)
D(z')
\int
\frac{k^2 dk}{2\pi^2}
P(k)
j_l (kr(z))
j_l (kr(z'))
\ea
$D,j_l$ are the growth function and spherical Bessel function, and the matter power spectrum $P(k)$ is defined as $\langle \delta_{{\bf k}} \delta^*_ { {\bf k'} }\rangle =(2\pi)^3 \delta( {\bf k} -{\bf k}') P (k)$ in terms of the matter fluctuation in $k$ space $\delta_{{\bf k}}$.

One would expect that the amplitude for the cross correlation among the different redshift bins would become smaller as the redshift difference becomes larger, which is verified in Fig. \ref{cell21z10} illustrating the the 21cm angular power spectra $C_l ^{21}$ at $z=10$. We take the frequency width $d \nu=1$MHz in our calculations (or equivalently the redshift bin width $dz=(1$MHz$/1.4 $GHz$)(1+z)^2$).
The 21cm angular power spectrum is proportional to the underlying matter power spectrum which has a peak at the scale corresponding to the matter-radiation equality. Hence, analogously to $C^{matter}_l$, $C_l^{21}$ also decreases for $l\gtrsim 100$.
$\overline{\Delta T_b}$ includes the integration of the mass function over the halo mass, and $\overline{\Delta T_b}$ becomes larger for a bigger abundance of minihalos. We also show the redshift dependence of $\overline{\Delta T_b}$ and $\bar{b}$ in Fig. \ref{figtbbias}. They have different redshift dependence and they are also different from the redshift dependence of $C_{l}^{matter}$ which evolves according to the growth function (proportional to the scale factor in the matter domination epoch $D(z)\propto R(z)$). We hence expect the 21cm tomographic data from multiple redshift bins would be of great help in the cosmological parameter estimation as discussed in the next section.
We also note that the axion isocurvature fluctuations enhance the small scale structures at a high redshift and the bias consequently is smaller for axion scenarios. There are less small structures for no axion scenarios at an earlier time, and hence the overdense regions are rarer and more biased.

Using the estimation of the 21cm fluctuation power spectra outlined in this section, we in the following sections attempt to find the range of $k_{osc}$ to which the forthcoming 21cm fluctuation signals are sensitive, followed by the discussions on the axion models relevant for such an observable $k_{osc}$ range.

\section{Forecasts}
\label{fisher}

We study how precisely the future 21cm fluctuation observations can constrain the axion properties, and we briefly outline the Fisher matrix analysis. 
The Fisher matrix for the angular power spectra is given by
\ba
F_{ij} = \sum _{\ell} \frac{2\ell+1}{2} f_\text{sky} \text{Tr} \left[  \frac{\partial {\boldsymbol C}_\ell}{\partial p_i} (\vec{p}) \widetilde{ {\boldsymbol C}}_\ell^{-1}(\vec{p})  \frac{\partial{{\boldsymbol C}_\ell}}{{\partial p_j}}(\vec{p}) \widetilde{{\boldsymbol  C }}_\ell^{-1}(\vec{p})
\right]
\ea
where $ \widetilde{ {\boldsymbol C}}_\ell =
{\boldsymbol C}_\ell+{\boldsymbol N}_\ell $ is the sum of signal and noise power spectra  \cite{Tegmark:1996bz,Knox:1995dq}. $\vec{p}$ is the vector consisting of the cosmological parameters $\{p_i\}$.

We estimate the noise power spectrum for the 21cm fluctuations, assuming the Gaussian beam window function, as \cite{Zaldarriaga:2003du,2006PhR...433..181F}
 \ba
N_l^{21}=\Delta_{21}^2 \theta_b^2 \exp\left(l(l+1) \frac{\theta_b^2}{8\ln 2}\right) 
\ea
where $\theta_b$ is the beam width 
and $\Delta_{21}$ is the telescope noise
\ba
\Delta_{21}=\frac{\lambda^2 }{A_{eff} \theta^2_b} \frac{T_{sys}}{\sqrt{\Delta \nu t} }
  \ea
  where $\lambda=21(1+z)$ cm, $A_{eff}$ is a total effective area, $t$ is the observation time 
  and $\Delta \nu$ is the bandwidth. 
    We included the effect of beam smearing by the Gaussian window function, so that the maximum multipole in our estimation scales as $l_{max}\propto 1/\theta_b$. The effective total area
represents the actually observed area for each $l$ mode,
$A_{eff}=(N_{dish} d^2l/l_{max}^2)A_{total}$
where $A_{total}$ is a geometrical total area
and $N_{dish}d^2 l$ represents the region of observed area in Fourier space.
The Fourier coverages of an interferometer are dependent on the array configuration and different among different $l$ modes.
It results in the $l$ dependence of $A_{eff}$.
 The exact array configuration however has not been set yet for the SKA and we simply follow Ref \cite{Zaldarriaga:2003du} by assuming that the antennas are distributed in a way to roughly realize the uniform Fourier coverage.$^4$\footnotetext[4]{Our $l$ independent $A_{eff}$ would be an overestimation (underestimation) of the noise for the red (blue) spectrum if the Fourier coverage can be simply parameterized as $d^2 l \propto (l/l_{max})^{n-1}$ ($n=1$ in our case and $n<1(n>1)$ is blue (red) spectrum). We leave more detailed array configuration dependent discussions for the future work \cite{McQuinn:2005hk}.}
We use $T_{sys}\sim T_{sky}\sim 180 (\nu/180$MHz)$^{-2.6}$ K in our estimation \cite{2006PhR...433..181F,Bowman:2008mk,Jelic:2008jg} indicating that the synchrotron emission foreground dominates the signal for a low frequency, and the redshift integration in our calculations is limited to the maximum redshift $z_{max}=20$ (as well as $z_{min}=6$ representing the end of reionization epoch).

 Analogously, for the CMB, $ N_l^{T,E}=(\theta_{FWHM} \Delta_{T,E})^2
       \exp
       \left[
         l(l+1)
         {\theta^2_{FWHM}}/{8\ln 2}
       \right]
       $
       with $\Delta$ representing the sensitivity of each frequency channel to the temperature/polarization and the corresponding noise power spectra for multiple channels are given by adding each channel contribution as $N_l=\left[    \sum_{\nu} N_{l,\nu}^{-1}  \right]^{-1}$ \cite{1987MNRAS.226..655B,Knox:1995dq}.

Our goal of this section is to find the $k_{osc}$ values which the 21cm signals from the minihalos can probe. The amplitude of the isocurvature perturbations is proportional to $k_{osc}^{-3}$ and too small values of $k_{osc}$ are already tightly bounded from the current large scale structure data. For instance, the CMB can give a bound  $k_{osc}\gtrsim 30$/Mpc and the current Ly$\alpha$ forest demands $k_{osc}\gtrsim 10^{3}$/Mpc \cite{Feix:2019lpo,Feix:2020txt,Irsic:2019iff}. 
Motivated by these current bounds, our study focuses on the relatively large $k_{osc} \gtrsim 10^3$/Mpc for which we currently lack the bounds and the forthcoming 21cm fluctuation observations can be a promising probe. We are, in particular, interested in how big a value of $k_{osc}$ can be measured by the 21cm signals and consequently the maximum value of the axion mass the future 21cm observations can probe. 
\begin{figure}[htbp]
  
   \begin{tabular}{c}

               \includegraphics[width=0.8\textwidth]{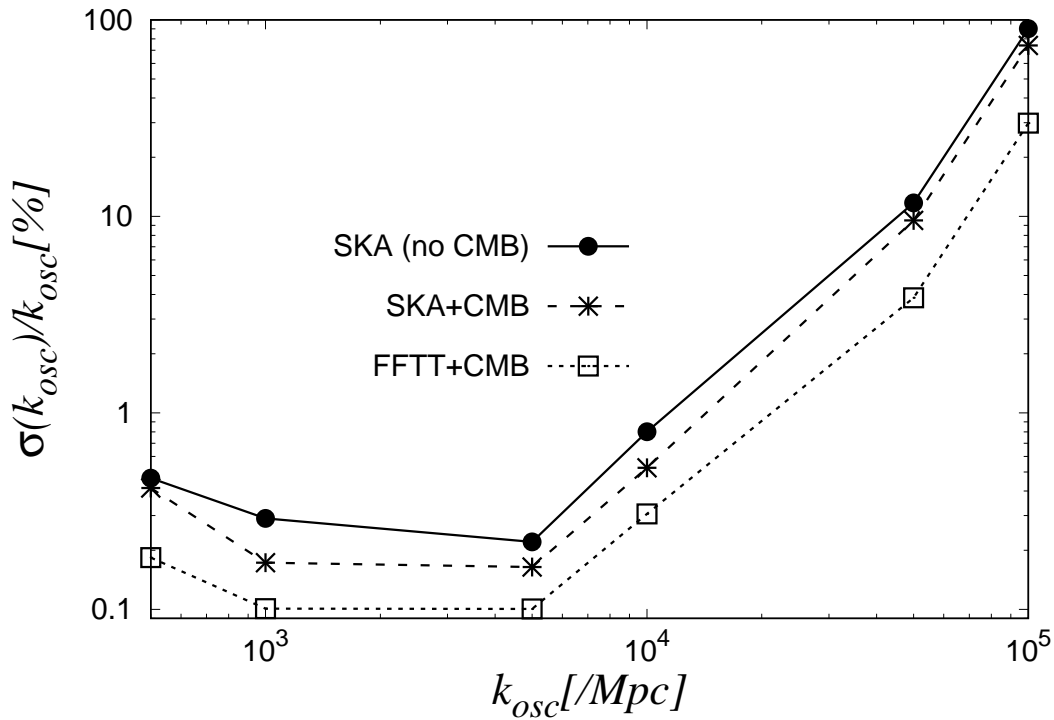}
    \end{tabular}

    
   \caption{
1$\sigma$ error $\sigma(k_{osc})/k_{osc}[\%]$ as a function of a fiducial value $k_{osc}[/$Mpc]. 
   }

   \label{figfisherfa1}
\end{figure}


\begin{figure}[htbp]
  
   \begin{tabular}{cc}
     \includegraphics[width=0.5\textwidth]{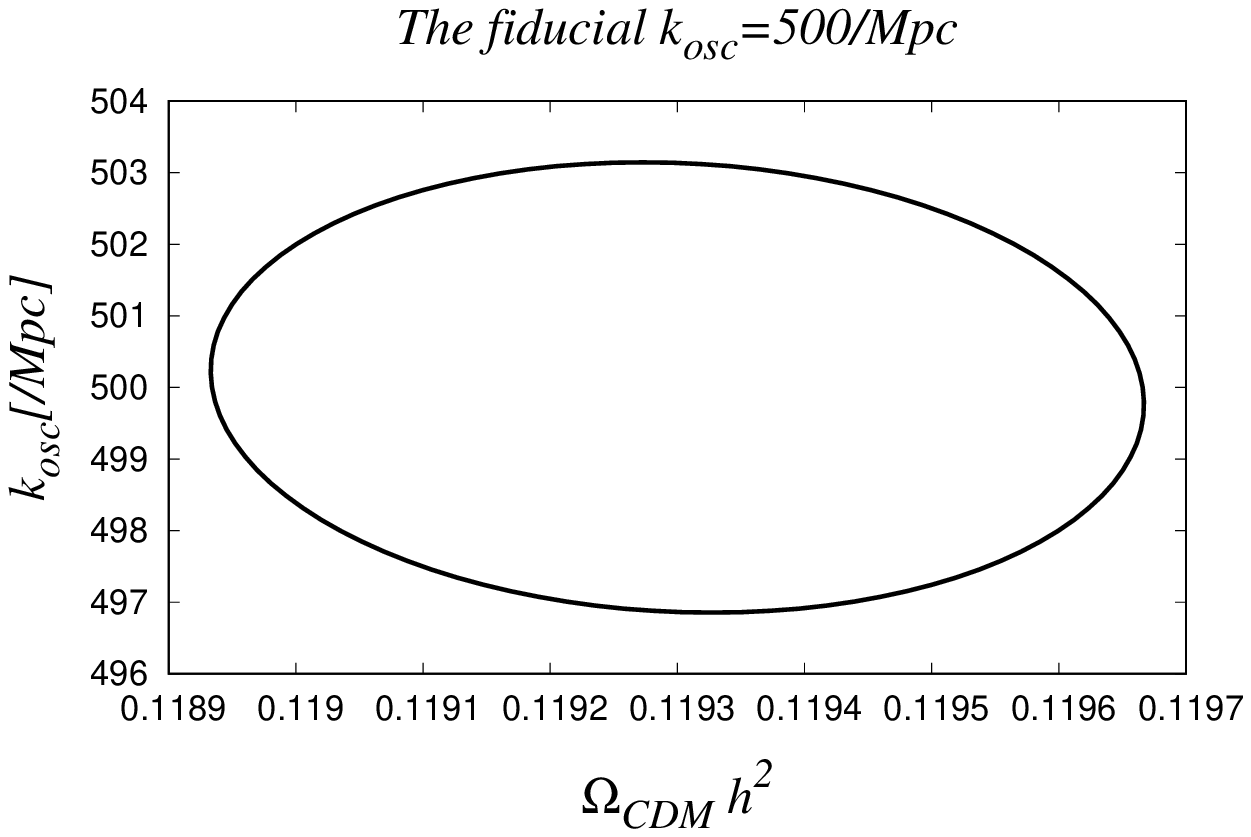}
     &
     \includegraphics[width=0.5\textwidth]{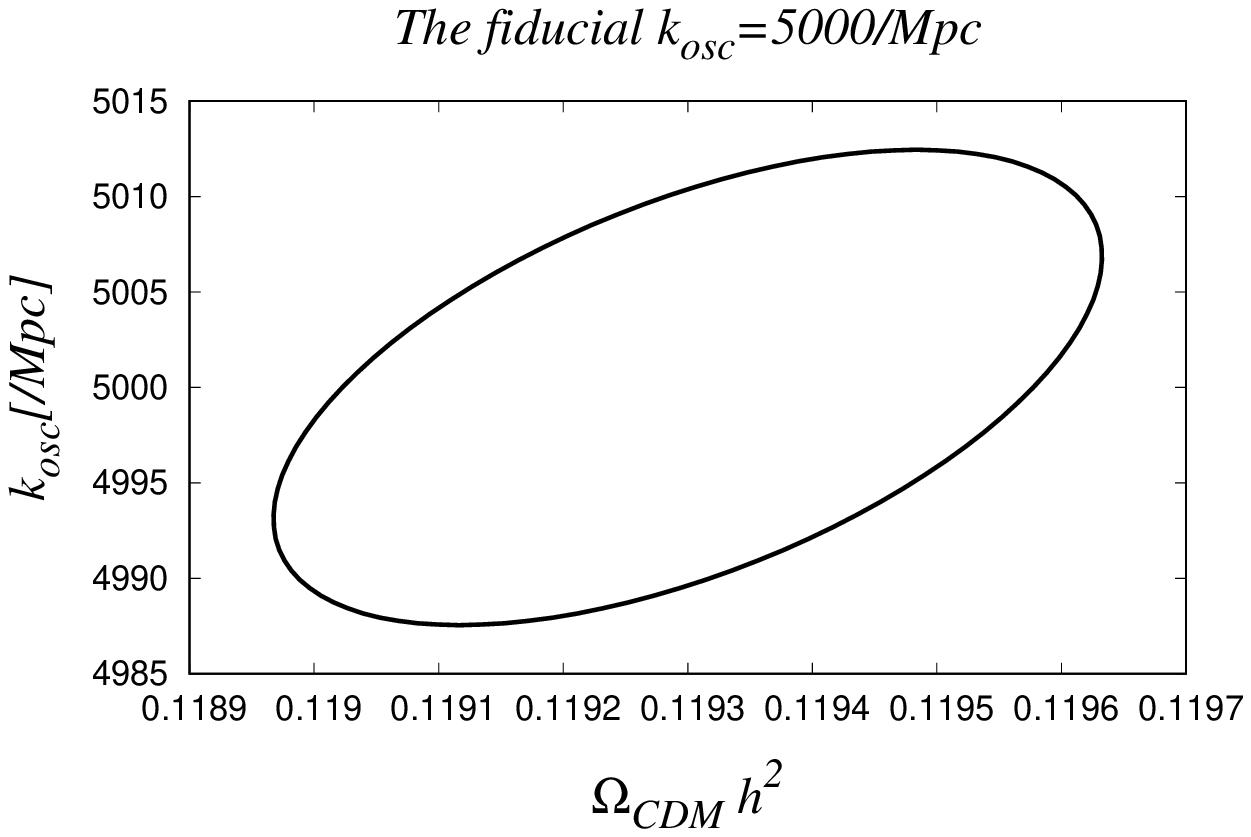}
    \end{tabular}

    
   \caption{Marginalized $1\sigma$ confidence ellipses for $\Omega_{CDM}h^2$ and $k_{osc}$[/Mpc]. The fiducial values are $(k_{osc},\Omega_{CDM}h^2)=(500$/Mpc$,0.11933)$ and $(5000$/Mpc$,0.11933)$. 
   }

 %


   \label{skaplkcont}
\end{figure}

\begin{figure}[htbp]
  
   \begin{tabular}{c}
     \includegraphics[width=0.8\textwidth]{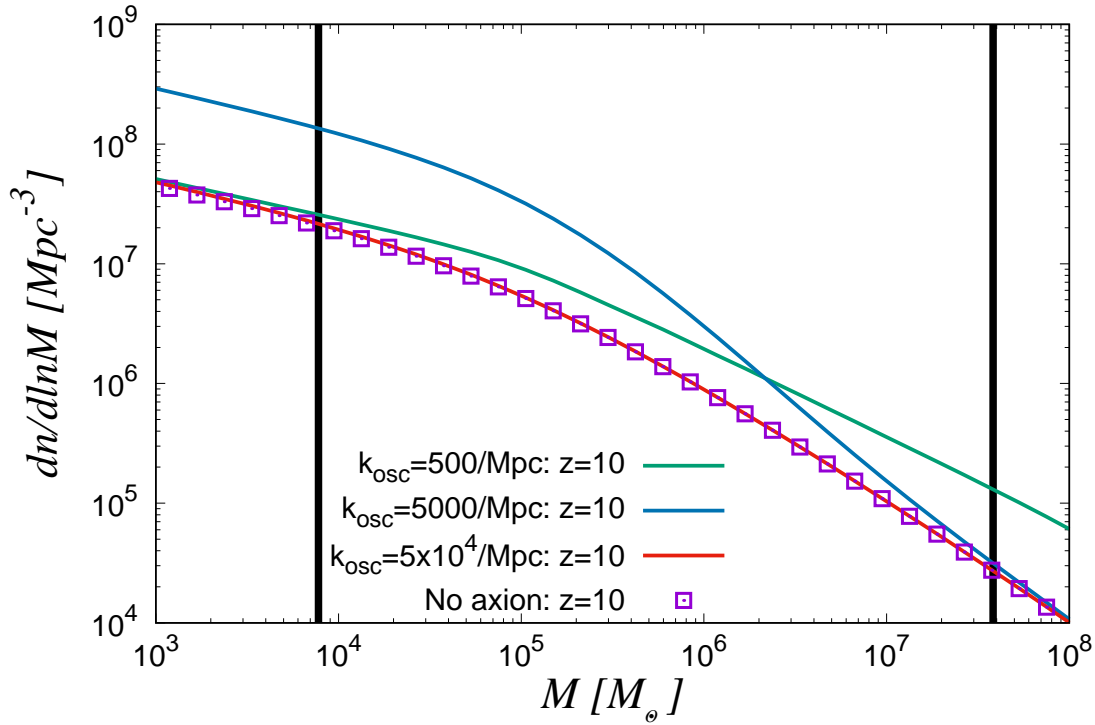}
    \end{tabular}

    
   \caption{The halo mass functions as a function of a halo mass. The vertical lines represent the minimal and maximal minihalo masses of our interest at $z=10$ (given by Eqs. (\ref{mminhalo}, \ref{mmaxhalo})). 
   }

 %
   \label{propermassfnz10k500}
\end{figure}
Fig. \ref{figfisherfa1} shows the $1\sigma$ errors for a given fiducial $k_{osc}$ value, where the other six $\Lambda CDM$ parameters (the fiducial values are CDM density $\Omega_c h^2=0.12,$ the baryon density $\Omega_b h^2=0.022,$ the reionization optical depth $\tau_{reion}=0.056,$ the spectral index and the amplitude of the adiabatic power spectrum $n_s=0.97, A_s=2.1\times 10^{-9}$ and the reduced Hubble parameter $h=0.7$) are marginalized over. The axion is assumed to constitute the whole dark matter of the Universe. 
For the 21cm, we assumed the SKA-like and FFTT(Fast Fourier Transform Telescope)-like specifications (the total effective area $A_{eff}$ [$m^2$], the band width $\Delta \nu$[MHz], the beam width $\Delta \theta$[arcmin], the integration time $t$ [hrs] are $(A_{eff},\Delta \nu, \Delta \theta,t)=$($10^5, 1, 9, 10^3$) for SKA and ($10^7, 1, 9, 10^3$) for FFTT) \cite{skawebpage,Tegmark:2008au}. For the CMB, we assumed the Planck-like specifications (the beam width (9.9,7.2,4.9) [arcmin], the temperature noise (31.3, 20.1, 28.5)[$\mu K$ arcmin] and the polarization noise (44.2, 33.3, 49.4)[$\mu K$ arcmin] respectively for the band frequency channels of (100, 147, 217)[GHz]) \cite{Planck:2006aa}.
Adding the CMB data improves the precision of $k_{osc}$ by lifting the parameter degeneracies, but we find the 21cm power spectra alone can also give the precise measurements because of the tomographic information from different redshifts. We also found the unmarginalized errors for $k_{osc}$ (i.e. fixing the other parameters besides $k_{osc}$) are smaller by more than an order of magnitude (for instance, for the fiducial value $k_{osc}=5 \times 10^3/$Mpc, the unmarginalized errors $\sigma(k_{osc})/k_{osc}=0.01\%$ and $0.005\%$ respectively for SKA and FFTT to be compared with the marginalized errors of $0.16\%$ and $0.10\%$). This can be expected because of the non-trivial parameter degeneracies, and the degeneracies between $\Omega_{CDM}h^2$ and $k_{osc}$ are illustrated in Fig. \ref{skaplkcont} assuming the SKA and CMB data. The other five parameters are marginalized over in these 1$\sigma$ error contours. While the decrease of minihalo abundance due to the variation of $k_{osc}$ can be compensated by the increase of $\Omega_{CDM}$, the slopes of the $1\sigma$ error contours in these figures change because of the non-trivial dependence of the mass function on $k_{osc}$ as illustrated in Fig. \ref{propermassfnz10k500} (where the relevant minihalo mass range (given in Eqs. (\ref{mminhalo},\ref{mmaxhalo})) are also indicated). The smaller $k_{osc}$ gives the larger isocurvature fluctuation amplitude $\propto k_{osc}^{-3}$ at a larger scale, and the abundance of a larger halos consequently could be enhanced. The larger $k_{osc}$ on the other hand can let the isocurvature perturbation with a cutoff scale $k_{osc}$ extend up to a smaller scale, so that the small halo formation can occur at an earlier epoch and the larger halo abundance consequently can be enhanced due to the those small halos' assembling to form the larger ones. We hence can expect there exists the optimal $k_{osc}$ which our 21cm signals from the minihalos can measure most precisely, and we found that our 21cm signals would be most sensitive to around $k_{osc}\sim 5\times 10^3/$Mpc as illustrated in Fig. \ref{figfisherfa1}. 
 We also found the 1$\sigma$ error $\sigma(k_{osc})/k_{osc}$ exceeds unity for $k_{osc}>1.2 \times 10^5/$Mpc for the SKA with the CMB, and for $k_{osc}>1.4 \times 10^5/$Mpc for the FFTT with the CMB.
We in the following attempt to find the maximum axion mass corresponding to $k_{osc}<1.2 \times 10^5/$Mpc as the axion mass range which can be probed by the forthcoming 21cm fluctuation observations.

\section{Axion mass probed by the 21cm fluctuations}
\label{axionmass}

\begin{figure} 
  \begin{center}
    \begin{tabular}{cc}

           \includegraphics[width=0.8\textwidth]{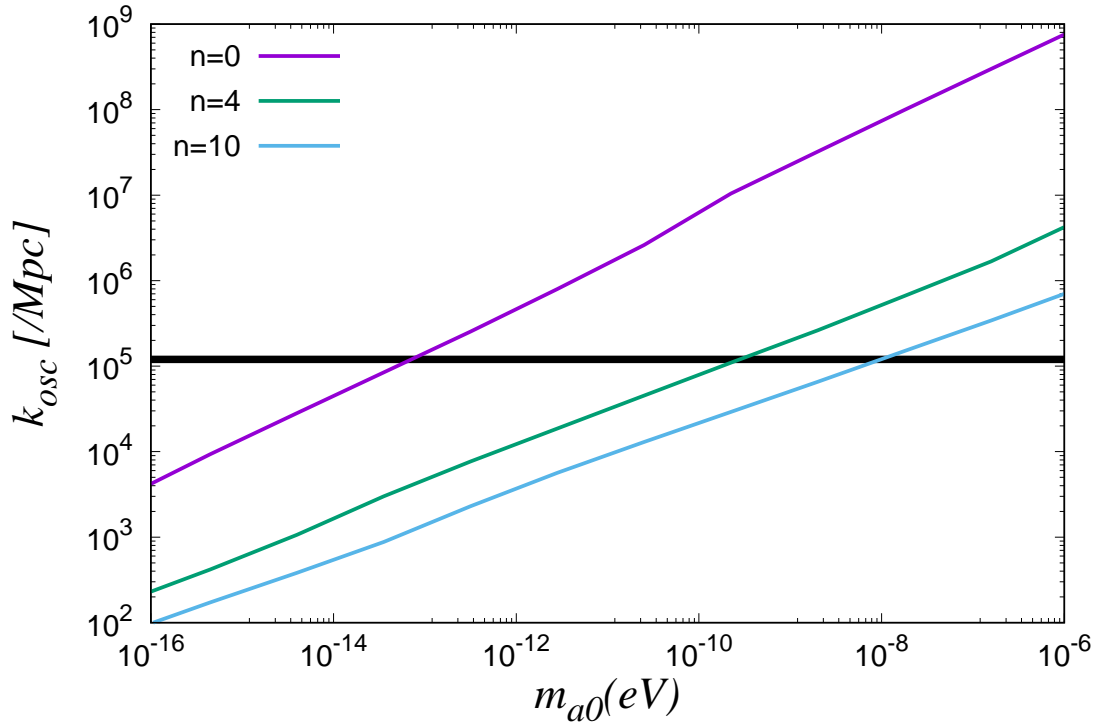}
    \end{tabular}
  \end{center}
  \caption{The comoving horizon scale $k_{osc}=R(T_{osc}) H(T_{osc})$ when the axion oscillation starts as a function of $m_{a0}$. $n$ represents the temperature dependence of the axion mass.
    The horizontal line indicates the maximum $k_{osc}=1.2\times 10^5$/Mpc below which $k_{osc}$ can be measured by the 21cm signals within 1$\sigma$ confidence level.
  }
  \label{ma0k0}
\end{figure}

We have found, from the Fisher matrix analysis in the last section, that the 21cm fluctuations from the minihalos can probe $k_{osc}< 1.2 \times 10^5/$Mpc, and we here discuss the axion parameters which can cover such a $k_{osc}$ range.
The temperature dependent axion mass is conventionally parameterized as $m_a=m_0 (T/\mu)^{-n}$ for $T>\mu$. $\mu$ represents a strong coupling scale which we parameterize as $\mu=\sqrt{m_{a0} f_a}$.$^5$\footnotetext[5]{$\mu$ is a model dependent parameter and could also depend on the UV completion and dark sector models (for instance, for the QCD axion model, $\mu\sim \Lambda_{QCD}\sim 2.5 \sqrt{m_a f_a}\sim 200$ MeV) \cite{fai2017,har2016}.} The axion mass $m_a(T<\mu)\equiv m_{a0}$ is temperature-independent for $T<\mu$ \cite{bo2016,Blinov:2019rhb,Borsanyi:2015cka}. We treat the temperature independent mass $m_{a0}$ as a free parameter and $f_a$ is determined from the requirement that all the dark matter consists of the axion. The model-dependent parameter $n$ represents the sensitivity of the axion mass on the temperature. The conventional QCD dilute instanton gas model, for instance, gives $n=4$ while the lattice QCD simulations give a slightly smaller value \cite{wan2009,dia2014,bo2016}. We simply treat $n$ as a free parameter in the range $n\leq 10$ when we relate $k_{osc}$ to $m_{a0}$.$^6$ \footnotetext[6]{See for instance Refs. \cite{har2016,fai2017,va2018,bus2019,Hui:2016ltb,Blinov:2019jqc,ari2012,Feix:2019lpo,Irsic:2019iff} discussing the axion cosmology for $n=0$ up to $n=20$.} $k_{osc}=R(T_{osc})H(T_{osc})$ ($R$ is a scale factor) is defined as the comoving horizon scale when the axion starts oscillation specified by the temperature satisfying $m_a(T_{osc})=3H(T_{osc})$. The oscillation starts deep in the radiation era for our scenarios because we are interested in the axion mass range $m_{a0}\gg  H(T_{eq})\sim 10^{-28}$eV. 
$k_{osc}$ is shown in Fig. \ref{ma0k0} as a function of $m_{a0}$, where we used Ref. \cite{hus2016} for the temperature evolution of the effectively massless degrees of freedom in the standard model. The horizontal line indicates our $1\sigma$ sensitivity from the SKA $k_{osc}<1.2\times 10^5$, which then means from this figure that the 21cm observations can probe axion mass up to $m_{a0}\sim 9.5 \times 10^{-14}$ eV for $n=0$. This can well exceed the maximum mass probed by the other means, notably the CMB which can have the sensitivity up to $m_{a0}\sim 10^{-20}$eV and the Ly$\alpha$ forest with the sensitivity up to $m_{a0}\sim 10^{-17}$eV \cite{Feix:2019lpo,Feix:2020txt,Irsic:2019iff}. Our study can be also complimentary to the 21cm forest observations which can probe up to the comparable mass scale $m_a\sim 10^{-12}$ eV \cite{Shimabukuro:2020tbs}.
  The axon mass range sensitive to the 21cm forest can be comparable to our studies because the 21cm forest observations also look at the signals from the minihalos, namely the 21cm absorption lines in the spectra emitted from the radio bright sources. The 21cm fluctuations could be however better than the 21cm forest for the precision on the axion parameter determination because of the better statistics by the tomographic information from many redshift bins, unless there are abundant observable radio loud sources at a high redshift for the 21cm forest observations. 
For a stronger temperature dependence, the detectable axion mass $m_{a0}$ can be even bigger. For $n=4$, it extends to $m_{a0}\sim 3.4\times 10^{-10}$ eV and, for $n=10$, we can probe up to $m_{a0}\sim 1.2 \times 10^{-8}$eV.

\section{Discussion/Conclusion}
\label{disconc}

We have so far assumed the total dark matter consists of the axion, and we here briefly mention the fractional axion dark matter scenarios. Those partial axion dark matter scenarios can be studied in analogy to the total axion dark matter scenarios by scaling the axion isocurvature perturbation power spectrum by a factor $(\Omega_a/\Omega_{CDM})^2$.

\begin{figure}[htbp]
  
   \begin{tabular}{c}

               \includegraphics[width=0.8\textwidth]{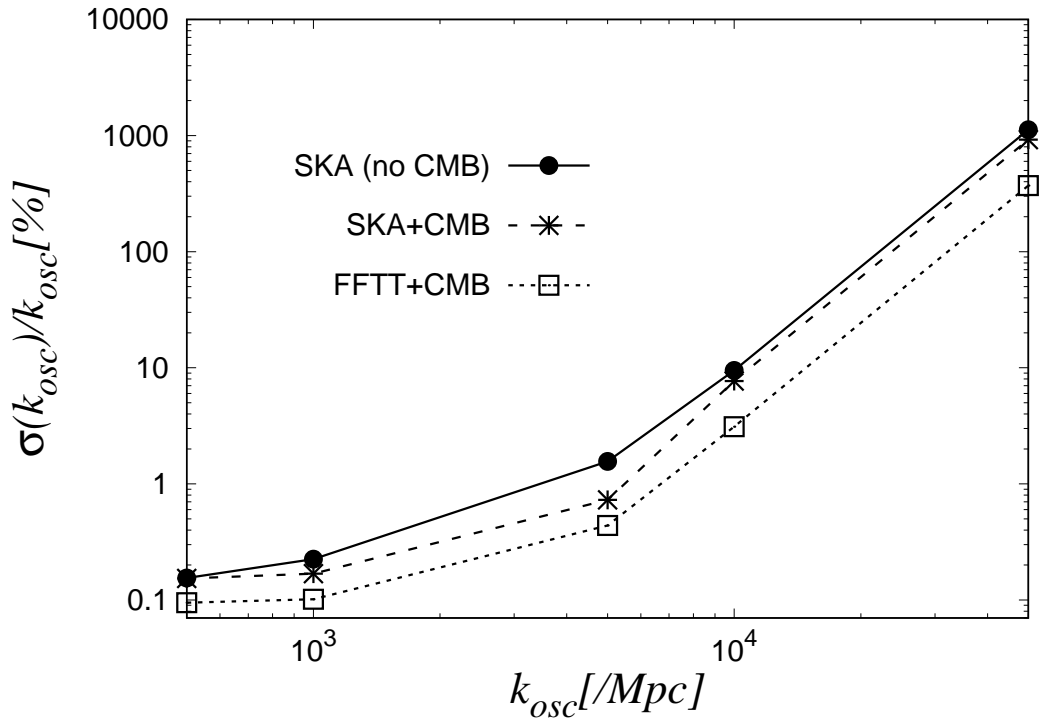}
    \end{tabular}

   \caption{
1$\sigma$ error $\sigma(k_{osc})/k_{osc}[\%]$ as a function of a fiducial value $k_{osc}[/$Mpc]. $\Omega_a/\Omega_{CDM}=0.1.$ 
   }

   \label{figfisherfa01}
\end{figure}

Fig. \ref{figfisherfa01} shows the marginalized 1$\sigma$ errors $\sigma(k_{osc})/k_{osc}$ (marginalized over the other six $\Lambda$CDM parameters) when the axion fraction is fixed to $10\%$ ($r_a\equiv \Omega_a/\Omega_{CDM}=0.1$). Because of the smaller isocurvature perturbation contributions compared with the total axion dark matter scenarios, the errors of $k_{osc}$ in general tend to become bigger.
There is a slight difference in the dependence of $\sigma(k_{osc})/k_{osc}$ on $k_{osc}$ for a relatively small $k_{osc}\sim 10^3$ between $r_a=1$ and $r_a=0.1$ scenarios. Such a different scale dependence is caused again by the evolution of the mass function which is affected by the amplitudes of isocurvature perturbations. The epoch when the small halo formation gets saturated (hence does not change the mass function slope anymore) for $r_a=1$ is earlier compared with the partial axion dark matter scenarios (the saturation occurs for the peak height $\nu=\delta_{crit}/\sigma(M)\ll 1$ so that the mass function possesses a conventional power law slope).
Such a non-trivial evolution of the mass function is reflected in the different scale dependence of the $k_{osc}$ error.

%

We note that the minihalos could be susceptible to the X-ray heating \cite{1997ApJ...475..429M}, and the bounds would be weakened in existence of the efficient IGM (intergalactic medium) heating because the baryon Jeans mass increases. While our analysis applies to the scenarios with a low IGM temperature and the current data still cannot exclude the relatively low IGM temperature scenarios ($T_{IGM}\sim {\cal O}(1\sim 10)$K at $z\sim 10$ depending on the unknown ionization state of the IGM \cite{2020MNRAS.493.4728G,2018ApJ...863..201A,2015ApJ...809...62P}), the gas temperature evolution history can be heavily model dependent (such as the properties of early heating sources) and we leave the more detailed analysis for future work including the scenarios with efficient gas heating.

Our analysis is applicable for the scenarios where the minihalo contributions to the 21cm fluctuations dominate those from the intergalactic medium (IGM). 
Such scenarios are indeed motivated and supported by the existence of axions because the minihalo contributions can be enhanced, while the IGM contributions do not get affected, thanks to the earlier formation and a bigger abundance of the minihalos than the conventional no-axion scenarios \cite{Furlanetto:2006xi}.
For instance, Fig. 2 indicates the axions can enhance the minihalo contributions about by a factor 10 even though the exact enhancement can be bigger or smaller depending on the choice of model parameters. 
We also note that the exact relative significance between the minihalo and IGM contributions can heavily depend on the IGM/reionization evolution histories, and it would be hard to completely exclude the possibility that the minihalos can dominate the IGM to constrain the axion parameters. 
For instance, the scenario illustrated by Ref. \cite{Sekiguchi:2017cdy} serves as an existence proof for a concrete scenario where the minihalo contributions can dominate the IGM ones and hence our analysis is applicable. 
Facing the lack of our knowledge on the early Universe histories and models, it would be worth exploring many possibilities to seek the hint on the new physics beyond the standard particle physics model.

More detailed noise analysis for a more realistic experimental setup beyond what has been done here would be also worth seeking. For instance the SKA's antenna distribution for which we simply assumed the uniform Fourier coverage could be modeled more properly and using the 3D $P(k)$ is also a possibility (see for instance \cite{McQuinn:2005hk}).
One advantage for using $P(k)$ is that the maps through the parallel components to the line-of-sight direction can be totally different from those through the transverse modes, which can be useful to deal with the noises and systematics. For instance, the continuum foreground emissions are typically smooth and power-law like, so that they are confined only to the lowest few modes of the parallel component. It may also be easy to understand $P(k)$ theoretically because $P(k)$ is directly related to the signal distributions in a real 3D volume. 

We also as a check performed the Fisher matrix analysis by worsening the noise levels by a factor 5 and 10 and the error on $k_{osc}$ changed by a factor a few. The results on the axion mass bounds consequently changed by just a factor a few, which is reasonable because the upper bounds on $k_{osc}$ come from the high $k_{osc}$ values where the sensitivity worsens abruptly regardless of the small change in the noise estimation. A factor of a few change in the axion mass parameter estimation can be easily migrated into other particle physics uncertainties such as the temperature dependence of the axion mass.

Our analysis in this paper is the first attempt to apply the 21cm fluctuations from the minihalos to the axions in the post PQ symmetry breaking scenarios, and the further studies including more detailed numerical simulations are warranted.

\begin{figure}[htb!]
  \begin{center}
    \begin{tabular}{c}
                  \includegraphics[width=0.8\textwidth]{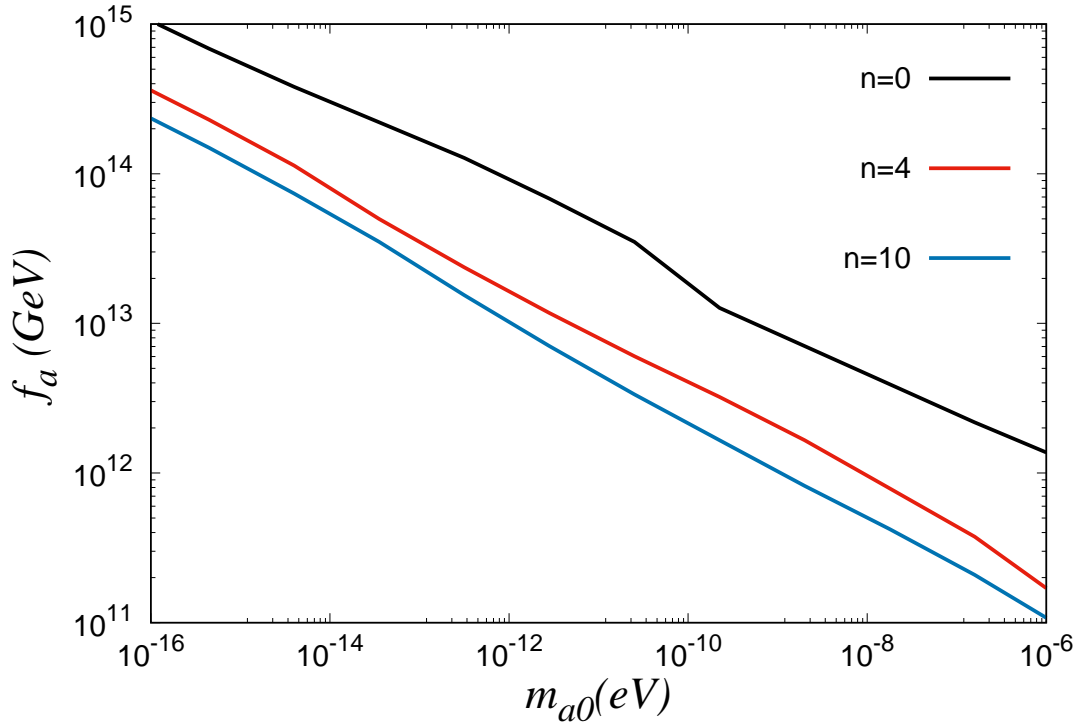}
    \end{tabular}
    %
  \end{center}
  \caption{The axion decay constant $f_a$ for $\Omega_a/\Omega_{CDM}=1$ as a function of the zero temperature axion mass $m_{a0}$.
  }
  \label{figtoscfa}
\end{figure}

Before concluding our discussions, let us show, for the consistency of the axion model parameters, the axion decay constant $f_a$ as a function of $m_{a0}$ in Fig. \ref{figtoscfa} which satisfies $\Omega_a h^2=0.12$ so that the axion makes up the whole dark matter of the Universe (the fractional axion dark matter scenarios can be studied by a simple scaling of $f_a$ because $f_a^2 \propto \Omega_a/\Omega_{CDM}$).

We estimated the current axion energy density as $\rho_a(T_{now})=m_{a0} n_a(T_{now})$, where the current axion number density is related to that when the axion oscillation starts by a scale factor $n_a(T_{now})=n_a(T_{osc}) (R(T_{osc})/R(T_{now}))^3$. $f_a$ can be obtained from $n_a(T_{osc})=\rho(T_{osc})/m_a(T_{osc})$ with $\rho_a(T_{osc})=m_a(T_{osc})^2 f_a^2 \theta^2/2$.

Let us briefly discuss the implication of these values of $f_a$ required for the consistent scenarios.



We assumed the post-inflation scenarios where the large isocurvature perturbations can arise due to the random axion field values in the causally disconnected Hubble patches. This assumption implies that the PQ symmetry breaking occurs after the inflation or, if it is broken during the inflation, the symmetry is restored after the inflation and broken afterwards. The former scenarios require the PQ symmetry breaking scale $f_a$ is at most the Gibbons-Hawking temperature scale during the inflation $T_{GH}=H_I/(2\pi)$ \cite{teg2008,ly1992} (the subscript `I' represents the inflation epoch), which leads to $f_a \lesssim 10^{13}$GeV assuming the slow-roll inflation with $r<0.07, A_s \sim 2 \times 10^{-9}$ \cite{ag2018}.   The latter scenarios with a symmetry restoration for a larger $f_a$ can be realized by the large thermal fluctuations due to the maximum temperature after the inflation exceeding $f_a$ and also by the non-thermal fluctuations due to the parametric resonance  \cite{chu1998,kof1995,kof1997,kol2003,giu2000,pal2004,kol1996jt,sh1994,tk1998,kasuya1997}.
For instance, the standard simple estimation of the maximum temperature which is attained during the reheating (this maximum temperature of the thermal bath can be much larger than the temperature at the beginning of the radiation-domination epoch $T_{RH}\sim (90/8\pi^3 g_*)^{1/4}$
referred to as the reheating temperature) is $T_{max}\sim  \sqrt{T_{RH} E_{I}}$ ($E_{I}$ represents the energy scale of inflation) \cite{kolb1990,pal2004,giu2000}. A large $f_a$ exceeding an aforementioned upper bound $10^{13}$ GeV can be possible for an efficient reheating process after the inflation even under the condition $E_{I} \lesssim 10^{16}$ GeV required from no observation of the primordial tensor modes \cite{ag2018}. We also add that the required value of $f_a$ for the desired axion dark matter abundance can be lower by an order unity factor by taking account of the anharmoinc effects, which is due to the deviation from the quadratic potential approximation delaying the onset of oscillations 
\cite{tu1985,ly1991,kob2013,fai2017}. The estimation of axion dark matter abundance can also be affected by an order unity factor due to the contribution of axions from the decay of topological defects, and the detailed numerical study of axion string-wall networks would be worth the further investigation \cite{kawa2014,hira2010,va2018,bus2019,hind2019,Armengaud:2019uso}. We leave the concrete inflation model building along with the exploration for the subsequent phenomenology for the future work.

In this paper, we studied the future prospects of the 21cm fluctuation observations on the axion dark matter in the post-inflationary PQ symmetry breaking scenarios. A key feature for such a scenario is the existence of the large axion isocurvature perturbations at smalls scale which is still allowed by the current data. The enhancement of the small-scale fluctuations can significantly affect the small scale structure formation, and we calculated the abundance of the small halos and demonstrated the 21cm emission signals from those minihalos can probe the axion mass up to $m_a\sim 10^{-13}$ eV for the temperature independent axion mass. The actual axion mass bound depends on the axion models (for instance we found the axion mass is detectable up to $m_a \sim  10^{-8}$eV for the temperature dependent axion mass ($n=10$ case in our discussions)), and the more realistic UV completion of the axion models as well as the inflation models for the consistent scenarios with a large axion decay constant would deserve the further study.
\\
\\This work was supported by the Institute for Basic Science (IBS-R018-D1) and Grants-in-Aid for Scientific Research from JSPS (17H01110, 18H04339).

\bibliography{/Users/kenji/GD/work/paper/kenjireference}


\end{document}